\documentclass[journal]{IEEEtran}
\usepackage[dvips]{graphicx}
\graphicspath{{./eps/}}
\DeclareGraphicsExtensions{.eps}
\usepackage{subfigure}
\usepackage{balance}
\usepackage{mathrsfs}
\usepackage{times}
\usepackage{bm}
\usepackage{subfigure}
\usepackage{color}
\usepackage{amsfonts}
\usepackage{amssymb}
\usepackage{graphicx}
\usepackage{algorithm}
\usepackage{algorithmic}
\usepackage{color}
\usepackage{multirow}
\usepackage{dsfont}
\usepackage{amsmath}
\usepackage{amsthm}
\usepackage{amsfonts}
\usepackage{amssymb}

\IEEEoverridecommandlockouts

\begin{document}

\title{Exploring Channel Estimation and Signal Detection for ODDM-based ISAC Systems}

\author{\IEEEauthorblockN{
{Dezhi Wang, Chongwen Huang, \IEEEmembership{Member,~IEEE}, Lei Liu, \IEEEmembership{Senior Member,~IEEE}, Xiaoming Chen, \\ \IEEEmembership{Senior Member,~IEEE}, Wei Wang, \IEEEmembership{Senior Member,~IEEE}, Zhaoyang Zhang, \IEEEmembership{Senior Member,~IEEE}, \\Chau Yuen, \IEEEmembership{Fellow,~IEEE}, and M{\'e}rouane  Debbah}, \IEEEmembership{Fellow,~IEEE}}\\ \vspace{-2em}
\IEEEcompsocitemizethanks{
	\IEEEcompsocthanksitem 
	D.~Wang, C. Huang, L. Liu, X. Chen, W. Wang, and Z. Zhang are with the College of Information Science and Electronic Engineering, and also with Zhejiang Provincial Key Laboratory of Information Processing, Communication and Networking, 
	Zhejiang University, Hangzhou 310027, China (E-mails: \{dz\_wang, chongwenhuang, lei\_liu, wangw, chen\_xiaoming, ning\_ming\}@zju.edu.cn).\protect
    \IEEEcompsocthanksitem C. Yuen is with the School of Electrical and Electronics Engineering, Nanyang Technological University, Singapore. (E-mail:
chau.yuen@ntu.edu.sg) \protect
    \IEEEcompsocthanksitem M. Debbah is with the Department of Electrical Engineering and Computer
Science and the KU 6G Center, Khalifa University, Abu Dhabi 127788, UAE,
and also with CentraleSupelec, University Paris-Saclay, 91192 Gif-sur-Yvette,
France (E-mail: Merouane.Debbah@ku.ac.ae).

	}
 }




\maketitle

\begin{abstract}
Inspired by providing reliable communications for high-mobility scenarios, in this letter, we investigate the channel estimation and signal detection in integrated sensing and communication~(ISAC) systems based on the orthogonal delay-Doppler multiplexing~(ODDM) modulation, which consists of a pulse-train that can achieve the orthogonality with respect to the resolution of the delay-Doppler~(DD) plane. To enhance the communication performance in the ODDM-based ISAC systems, we first propose a low-complexity approximation algorithm for channel estimation, which addresses the challenge of the high complexity from high resolution in the ODDM modulation, and achieves performance close to that of the maximum likelihood estimator scheme. Then, we employ the orthogonal approximate message-passing scheme to detect the symbols in the communication process based on the estimated channel information. Finally, simulation results show that the detection performance of ODDM is better than other multi-carrier modulation schemes. Specifically, the ODDM outperforms the orthogonal time frequency space scheme by 2.3 dB when the bit error ratio is $10^{-6}$.
\end{abstract}

\section{Introduction}
With the emergence of many high-mobility scenarios, such as satellites, unmanned aerial vehicles, and the Internet of vehicles, future 6G networks are required to provide high-quality communications for various applications~\cite{fei2023air}. However, traditional orthogonal frequency division multiplexing~(OFDM) struggles to ensure reliable communications in such high-mobility scenarios, primarily due to the severe Doppler effect~\cite{hwang2008ofdm,Zhang2023radar}.  

In response to these challenges, orthogonal time frequency space~(OTFS) has been developed and is now regarded as a promising communication paradigm to conquer the inter-carrier-interference~\cite{wang2021pilot}. In particular, OTFS effectively mitigates the influence of Doppler shift, offering strong resilience and excellent performance in fast time-varying channels~\cite{li2023pulse}.  In the process of OTFS modulation, the signal in the delay-Doppler~(DD) domain is transformed and placed onto the time frequency~(TF) domain by leveraging the inverse symplectic finite Fourier transform~(ISFFT). It is then transmitted using a TF plane rectangular pulse that maintains the orthogonality. Therefore, the OTFS is essentially a precoded OFDM. However, the ideal pulse in the TF plane cannot be obtained in practice~\cite{lin2023multi}. To achieve a realizable DD plane orthogonal pulse, Lin \emph{et. al}~\cite{lin2022orthogonal} proposed a novel multi-carrier~(MC) modulation on the delay-Doppler plane, named orthogonal delay-Doppler division multiplexing~(ODDM) modulation, which is an MC modulation in the DD domain and can avoid the shortcomings of OTFS mentioned above. The proposed advanced transmission pulse is designed as a pulse-train that maintains orthogonality with respect to~(w.r.t.) the resolution of the DD plane. This characteristic significantly enhances the potential of ODDM signals for facilitating integrated sensing and communication~(ISAC) applications~\cite{lin2023multi}. Moreover, in the frame of ISAC, the channel state information is significant for improving the communication performance~\cite{gan2021ris,channel2021li}. As a result, one can naturally integrate the ODDM into high-mobility ISAC scenarios to enhance communication performance by exploiting the channel information. 

In recent years, there have been existing works on  OTFS-based ISAC systems. Specifically,  Yang~\emph{et. al}~\cite{yang2024sensing} focused on the sensing-aided uplink transmission in an ISAC vehicular network based on the OTFS modulation by joint consideration of the parameter association, channel estimation, and signal detection. 
Yuan~\emph{et. al}~\cite{yuan2021integrated} explored the use of the ISAC technique for assisting OTFS transmission in both the uplink and downlink of vehicular communication systems. In addition, Priya~\emph{et. al}~\cite{priya2023channel} proposed a receiver design for the mmWave OTFS system, which includes channel estimation and data detection. Although, the existing work has demonstrated the effectiveness of the OTFS modulation in high-mobility scenarios, how to integrate the advanced ODDM scheme in the ISAC system hasn't been well discussed in the existing literature. Moreover, existing modulation methods cannot enable more precise and reliable ISAC functionalities.

Motivated by the above discussion, in this letter, we investigate the new ODDM modulation for the ISAC system in a high-mobility network, which enhances communication reliability by ensuring orthogonality with the DD plane resolutions. Specifically, we first estimate the channel parameters by designing a low-complexity approximation algorithm to obtain similar results earned by the maximum likelihood estimator~(MLE) method in the sensing process, which overcomes the high complexity caused by the large DD plane in the ODDM modulation. On acquiring the channel parameters, we utilize the orthogonal approximate message passing~(OAMP) to detect the symbols within the communications process, which achieves 
Bayesian optimality.
Finally, we validate the efficiency of the ODDM modulation through simulations. The results demonstrate that the detection performance of ODDM outperforms the OTFS by 2.3 dB when the bit error ratio~(BER) is $10^{-6}$, which validates the performance of the proposed algorithm.

The rest of the letter are organized as follows: Section II introduces the system model, while Section III introduces the proposed algorithm for the ISAC system based on the ODDM. In Section IV, we present the simulation results. Finally, we conclude the letter in Section V. 


\section{System Model}
In this paper, we consider the ISAC system with a transmitter and a target receiver, where the transmitter conveys the message while estimating the related channel parameters from the reflected signal. For simplicity, the transmitter and receiver are equipped with a single antenna\footnote{The scenario can be extended to multiple antennas by adding the steering vector.}.  The channel between the transmitter and receiver has multiple paths. The ISAC system utilized the time-division multiple access~(TDMA) to execute the sensing and communication functions. At first, the transmitter transmits the sensing signal to the receiver, and then it collects the sensing echoes for estimation and transmits the communication symbols to the receiver. We will proceed to present the physical model, along with the ODDM modulation and demodulation in the following. In this model, the total bandwidth is set as $B$ and the carrier frequency is $f_c$. 
\subsection{Physical Model}
Similar to the OFDM, ODDM is also a multi-carrier~(MC) modulation, therefore, we divide the bandwidth $B$ into $N$ subcarriers and the bandwidth of each subcarrier is $\Delta f$, i.e., $B=N\Delta f$. Define $P$ as the number of paths of the channel, thus the time-frequency~(TF) channel $h(t,\tau)$ is expressed as 
\begin{eqnarray}\label{channel model}
h(t,\tau)=\sum_{p=0}^{P-1}h_p\delta(\tau-\tau_p)e^{j2\pi\nu_pt},
\end{eqnarray} 
where $h_p\in\mathbb{C}$, $\tau_p=\frac{r_p}{c}$, and $\nu_p=\frac{v_pf_c}{c}$  represent the complex channel gain, the one-way delay and Doppler shift of $p$th path, respectively. $v_p$ and $r_p$ are the corresponding speed, distance of the $p$th path, and $c$ denotes the speed of light.

Considering the delay and Doppler effects that accompany  each path of the channel for the transmitted signal $s(t)$, we can calculate the received signal $y(t)$ as follows:
\begin{eqnarray}
y(t)=\sum_{p=0}^{P-1}h_ps(t-\tau_p)e^{j2\pi v_p(t-\tau_p)}+n_s(t),
\end{eqnarray}
where $n_s(t)$ is the corresponding additional white Gaussian noise~(AWGN).
\subsection{ODDM Modulation and Demodulation}\label{section_ODDM_modulation}
In this part, we present the process of ODDM modulation and demodulation, which is a type of  MC modulation in the DD domain. Within this ODDM modulation, the information data is assigned to a set of $M\times N$ data symbols in the DD domain, with $M$ and $N$ representing the number of time slots and subcarriers, respectively. Define $S(m,n)$ as the transmission data symbol, where $m\in\{0,1,\cdots,M-1\}$, $n\in\{0,1,\cdots,N-1\}$ represents the delay index and Doppler index, respectively. This implies that we have $MN$ data symbols that need to be modulated. Each MC symbol is then combined with $N$ subcarriers. 
Additionally,  we define $T$ as the duration of each time slot. In the DD domain, the spacing between subcarriers is $\frac{1}{NT}$, indicating that the symbol period is $NT$. The MC symbols are spaced by a short interval of $\frac{T}{M}$, which corresponds to the delay resolution. This staggered signal structure is a result of this spacing. The modulation process of ODDM is shown in Fig.~\ref{fig:ODDM_modulation}, which can be expressed as
\begin{eqnarray}
s(t)=\sum_{m=0}^{M-1}\sum_{n=0}^{N-1}S(m,n)u\Big(t-m
	\frac{T}{M}\Big)e^{j2\pi\frac{n}{NT}\big(t-m
		\frac{T}{M}\big)},
\end{eqnarray}
where $u(t)=\sum_{\hat{n}=0}^{N-1}a(t-\hat{n}T)$. Here, $a(t)$ satisfies a real-valued square-root Nyquist pulse that is time-symmetric. This has a time duration of $2Q\frac{T}{M}$ and complies with the following conditions:
\begin{figure}
	{	\centering\includegraphics[width=3.3in]{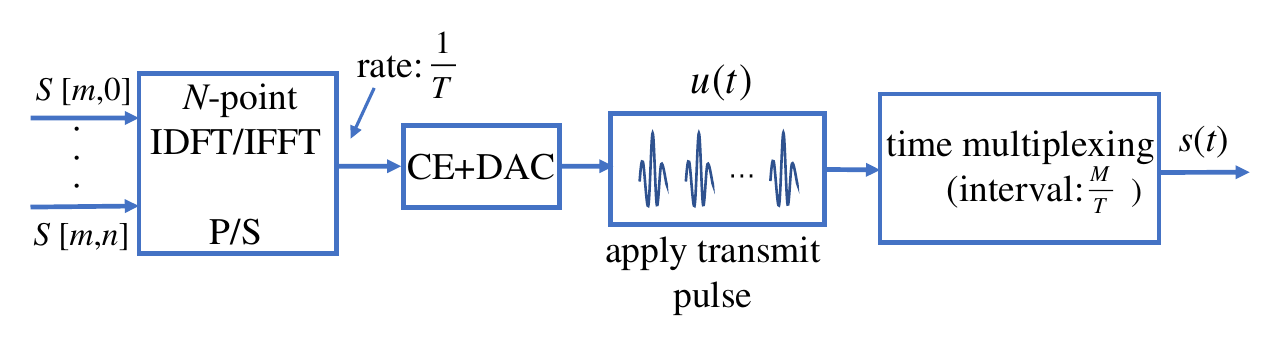}
		\caption{ODDM modulation}
\label{fig:ODDM_modulation}
	}
\end{figure}
\begin{eqnarray}
    \int_{-\infty}^{\infty}|a(t)|^2dt=\frac{1}{N},
\end{eqnarray}
where $Q$ is an integer and $Q\ll M/2$. As a result, we have $\int_{-\infty}^{\infty}|u(t)|^2dt=1.$
When $2Q\ll M$, $u(t)$ is orthogonal w.r.t. the delay and Doppler resolutions\footnote{The related proof can be seen in~\cite{lin2022orthogonal}.}, which is different from the rectangular waveform in the OTFS. 

Then, we give the received signal at the receiver end after passing the physical model in \eqref{channel model} as 
\begin{eqnarray}
&&y(t)=\sum_{p=0}^{P-1}h_ps(t-\tau_p)e^{j2\pi v_p(t-\tau_p)}+n_s(t)\nonumber\\
&&~~~~~
=\sum_{p=0}^{P-1}\sum_{m=0}^{M-1}\sum_{N=0}^{N-1}h_pS(m,n)u\Big(t-(m+l_p)\frac{T}{M}\Big)\nonumber\\
	&&~~~~~~\times e^{j2\pi\big(\frac{n+k_p}{NT}(t-(m+l_p)T/M)\big)}e^{j2\pi\frac{k_pm}{MN}}+n_s(t),
\end{eqnarray}
where $l_p=\tau_p\frac{M}{T}$ and $k_p=\nu_p NT$ denote the delay index and the Doppler index, respectively. Leveraging the orthogonality of $u(t)$,  we deduce the received ODDM signal in the DD domain subsequent to the matched filter based on $u(t)$ in the following manner~\cite{lin2022multicarrier} :
\begin{eqnarray}\label{sensing_model_received_signal_DD_domain}
Y(m,n)=\sum_{p=0}^{P-1}h_p\hat{S}(\hat{m},\hat{n})e^{j2\pi\frac{k_p(m-l_p)}{MN}}+n_s(m,n),
\end{eqnarray}
where $\hat{S}(\hat{m},\hat{n})=S(m,n)$ when $\hat{m}=m-l_p\geq0$ and $\hat{n}=[n-k_p]_N$, where $[\cdot]_N$ is the mod $N$ operator, and $n_s(m,n)$ is the noise signal sample in the DD domain. When $\hat{m}=m-l_p<0$, we have 
\begin{eqnarray}
\hat{S}(\hat{m},\hat{n})=e^{-j2\pi\frac{\hat{n}}{N}}S(M+\hat{m},\hat{n}).
\end{eqnarray}
\section{Channel Estimation and Signal Detection}
In this section, we give the details about the proposed algorithm for channel estimation and signal detection. First, we derive the effective channel matrix for ODDM modulation. Then we estimate the channel parameters by proposing a low-complexity iterative algorithm in the sensing process. Finally, the OAMP scheme is employed to detect the signal in the communication process based on the estimated parameters.
\subsection{Derivation of the Effective Channel Matrix}

Based on the received signal in~\eqref{sensing_model_received_signal_DD_domain}, we have the input-output relationship as follows~\cite{lin2022multicarrier}:
\begin{eqnarray}\label{effective input-output}
\bm{y}=\bm{H}\bm{s}+\bm{n}_s,
\end{eqnarray}            
where $\bm{y}\in\mathbb{C}^{NM\times1},\bm{s}\in\mathbb{C}^{NM\times1}$ denotes the vectorized form of transmit and receive signal in the DD domain, respectively. $\bm{n}_s\in\mathbb{C}^{NM\times1}$ with the zero mean and covariance $\sigma_w^2\bm{I}_{MN}$ is the corresponding noise vector, and $\bm{H}\in\mathbb{C}^{MN\times MN}$ is the effective channel matrix in the DD domain, which is expressed as
\begin{align}
\bm{H} = \resizebox{2.8in}{!}{$
\begin{bmatrix}
	\bm{A}_0^0 & & & \bm{A}_{L-1}^0\bm{D} & \cdots & \bm{A}_1^0\bm{D} \\
	\vdots & \ddots & & & \ddots & \vdots \\
	\bm{A}_{L-2}^{L-2} & \ddots & \bm{A}_0^{L-2} & \bm{0} & & \bm{A}_{L-1}^{L-2}\bm{D} \\
	\bm{A}_{L-1}^{L-1} & \ddots & \ddots & \bm{A}_0^{L-1} & & \\
	& \ddots & \ddots & \ddots & \ddots & \\
	\bm{0} & & \bm{A}_{L-1}^{M-1} & \cdots & \cdots & \bm{A}_0^{M-1}
\end{bmatrix}$}
\end{align}
where $\bm{D}=\text{diag}\left\{1,e^{-j\frac{2\pi}{N}},\cdots,e^{-j\frac{2\pi(N-1)}{N}}\right\}$ is additional phase rotation term, $L_1$ and $L$ represent the maximum delay index and Doppler index, and 
\begin{eqnarray}\label{element of effective matrix}
	\bm{A}_l^m=\sum_{l_1=-L_1}^{L_1}g(l_1+L_1+1,l)e^{j2\pi\frac{l_1(m-l)}{MN}}\bm{C}^{l_1},
\end{eqnarray}
where $\bm{C}\in\mathbb{R}^{N\times N}$ is the cyclic permutation matrix 
and $g(l_1+L_1+1,l)$  is the element of $(2L_1+1)\times L$ DD domain channel matrix $\bm{G}$. The Doppler and delay of $g(l_1+L_1+1,l)$ are $l_1\frac{1}{NT}$ and  $l\frac{T}{M}$, respectively. The non-zero elements in $\bm{G}$ reflect the channel gain, which means that non-zero elements of the matrix $\bm{G}$ are equal to the number of channel paths.

Since $g(l_1+L_1+1,l)$ represents the channel gain $h_p$ of 
$p$th path, $\bm{A}_l^m$ can be expressed by the channel information of $P$ paths based on~\eqref{element of effective matrix}. Consequently, the non-zero elements of the effective channel matrix $\bm{H}$ in~\eqref{effective_matrix} can be decomposed~as 
\begin{eqnarray}
 \bm{H}=\sum_{p=0}^{P-1}h_p\bm{H}_p,   
\end{eqnarray}
where $\bm{H}_p\in\mathbb{C}^{MN\times MN}$ represents the coefficient matrix related to the path $p$. Therefore, the input-output relationship in~\eqref{effective input-output} can be rewritten as 
\begin{eqnarray}\label{equivalent_equation}
	\bm{y}=\sum_{p=0}^{P-1}h_p\bm{H}_p\bm{s}+\bm{n}_s.
\end{eqnarray}

\subsection{Channel Estimation in the Sensing Process}
As shown in~\eqref{channel model}, the channel parameters $\tilde{\bm{H}}_p$ in the sensing process consist of the complex channel gain $h_p$, the delay $\tau_p$, and the Doppler $\nu_p$. In this part, we estimate the channel parameters by approximating the solution of the MLE method using an approximate iterative scheme, which can reduce the complexity. At first, we define $\bar{\bm{\Lambda}}=(\bar{h}_0,\cdots,\bar{h}_{P-1},\bar{\tau}_0,\cdots,\bar{\tau}_{P-1},\bar{\nu}_0,\cdots,\bar{\nu}_{P-1} )$ as the true value of the unknown parameters. In the sensing process,  the transmission symbol $\bm{s}$ is known in advance, thus the parameters estimation problem is expressed as
\begin{eqnarray}\label{MLE}
\hat{\bm{\Lambda}}=\arg\min_{\bm{\Lambda}}l(\bm{y}|\bm{\Lambda},\bm{s}),
\end{eqnarray}
where $l(\bm{y}|\bm{\Lambda},\bm{s})=|\bm{y}-\sum_{p=0}^{P-1}h_p\tilde{\bm{H}}_p\bm{s}|^2$ is the corresponding ML estimator.

It is unrealistic to utilize the brute-force search method to solve the minimum problem, thus we propose a low-complexity method to approximate the solution to the problem. Since $l(\bm{y}|\bm{\Lambda},\bm{s})$ is quadratic w.r.t. $h_p$ for fixed $\{\tau_p,\nu_p\}$, when $\{\tau_p,\nu_p\}$ are given, the solution to~\eqref{MLE} w.r.t. $h_p$ satisfies the following equations
\begin{eqnarray}\label{mid_equation}
	\sum_{p'=0}^{P-1}h_{p'}\bm{s}^H\tilde{\bm{H}}_p\tilde{\bm{H}}_{p'}\bm{s}=\bm{s}^H\tilde{\bm{H}}_p^H\bm{y},  \forall p\in\{0,\cdots, P-1\},
\end{eqnarray}
Therefore, we can obtain $\{h_p, p=0,\cdots, P-1\}$ by solving the linear system of equations, which contains $P$ linear equations and can be solved by some methods, such as the Gaussian elimination method. Based on~\eqref{mid_equation}, the minimization problem w.r.t. $\{\tau_p,\nu_p\}$ in~\eqref{MLE} can be transformed into the following maximization problem
\begin{eqnarray}\label{maximum_problem}
&&\{\hat{\bm{\tau}},\hat{\bm{\nu}}\}=\arg\max_{\bm{\tau},\bm{\nu}}l_1(\bm{y}|\bm{\Lambda},\bm{s})\nonumber\\
&&~~~~=\arg\max_{\bm{\tau},\bm{\nu}}\sum_{p=0}^{P-1}\bigg\{\frac{|\bm{s}^H\tilde{\bm{H}}_p\bm{y}|}{\bm{s}^{H}\tilde{\bm{H}}_p^H\tilde{\bm{H}}_p\bm{s}}\nonumber\\
&&~~~~~~~~~-\frac{\sum_{p'\neq p}h_{p'}\bm{s}^{H}\tilde{\bm{H}}_p^H\tilde{\bm{H}}_{p'}\bm{s}\bm{y}^H\tilde{\bm{H}}_p\bm{s}}{\bm{s}^{H}\tilde{\bm{H}}_p^H\tilde{\bm{H}}_p\bm{s}}\bigg\},
\end{eqnarray}
where the first term and the second term of the right-hand side of~\eqref{maximum_problem}, denoted by $Q_p(\tau_p,\nu_p)$ and $I_p(\{h_{p'}\}_{p'\neq p},\bm{\tau},\bm{\nu})$, respectively, are the useful signal and the interference for path $p$, respectively. In the sensing process, the channel gain $h_p$ is unknown. As a result, $\hat{\tau}_p,\hat{\nu}_p$ cannot be obtained by solving the maximization problem $l_1(\bm{y}|\bm{\Lambda},\bm{s})$. Considering the sparse characteristic of the matrix $\tilde{\bm{H}}_p$, the dependence of the interference term $I_p$ on all terms $\{\tau_{p'},\nu_{p'}\}$, $p'\neq p$ tends to weak~\cite{gaudio2020effectiveness}. Consequently, we utilized a block-wise optimization algorithm to update the pair $\{\tau_p, \nu_p\}$ and channel gain $h_p$ alternately. The detail of the algorithm is shown in Algorithm~1.
\begin{algorithm}[htbp]
	\caption{Low-complexity Algorithm  for Channel Estimation}\label{algorithm_PDHG}
	\begin{algorithmic}[1]
		\STATE  \textbf{Initialize}: iteration $k=0$, $h_p(0),\tau_p(0),\nu_p(0),~\forall p\in\{0,\cdots, P-1\}$, thresholds $\varepsilon$.
		\WHILE{$n \leq Itera_{\max}$}   
		\STATE update $\{\tau_p(n),\nu_p(n)\},~\forall p\in\{0,\cdots, P-1\}$ by solving $\arg\max_{\tau_p,\nu_p}\{Q_p(\tau_p,\nu_p)-I_p(\{h_{p'}\}_{p'\neq p},\tau_p,\nu_p,\tau_{p'},\nu_{p'})\}$;
		\STATE update the channel gain $h_p(n),~\forall p\in\{0,\cdots, P-1\}$ by solving \eqref{mid_equation};
		\IF{$|h_p(n)-h_p(n-1)|+|\tau_p(n)-\tau_p(n-1)|+|\nu_p(n)-\nu_p(n-1)|\leq \varepsilon,~\forall p\in\{0,\cdots,P-1\}$}
		\STATE break
		\ENDIF
		\ENDWHILE 
		\RETURN Channel gain, delay, and Doppler $h_p,\tau_p,\nu_p,~\forall p\in\{0,\cdots, P-1\}$
	\end{algorithmic}
\end{algorithm}
In the practical ODDM modulation, we maximize the problem by pursuing the integer delay-Doppler grid, where the channel effective matrix has been pre-computed and saved. The dual-step maximization search method can reduce the overall computational complexity, which is suitable for larger block dimension $MN$ in the ODDM modulation.

As introduced in the introduction, the ODDM modulation is robust, which is proved in the simulation. In addition, the proposed algorithm can estimate channel information quickly due to its low complexity. The delay and Doppler shift can be seen unchanged over a short slot. In this scenario, we only estimate the path gain, which is a linear system of equations and easy to obtain. 
\subsection{Symbol Detection in the Communication Process}
The estimated channel parameters in the sensing process can be employed to improve the performance of symbol detection in the communication process. As we have implemented the TDMA model for sensing and communication, the sensing and communication channels are identical~\cite{liu2020uplink}, therefore we define $\bm{H}$ as the channel information. Based on the sparsity of the effective channel in~\eqref{effective_matrix}, we utilize the OAMP to detect the symbols in the communication process based on the estimated channel information, which can relax the restrictions on the channel matrix~\cite{liu2023oamp} and solve the sparse linear inverse problem, such as~\eqref{effective input-output}. The core substance of the OAMP method centers on two estimators, i.e., ``de-correlated" linear estimator~(LE) and ``divergence-free"  non-linear estimator~(NLE), where LE predominantly serves to mitigate the impacts of multi-path interference and Doppler shift, while the NLE plays an integral role in curbing the Gaussian noise and improve the quality of the estimate of $\hat{\bm{s}}$~\cite{cheng2020integral}.  The two estimators are summarized as 
\begin{align}
&\text{LE}: \bm{r}_t=\bm{s}_t+\frac{1}{\epsilon_t^\phi}\bm{H}^H\Big(\bm{H}\bm{H}^H+\xi_t\bm{I}\Big)^{-1}(\bm{y}-\bm{Hs}_t),\label{LE} \\
&\text{NLE}: \bm{s}_{t+1}=\eta_t(\bm{r}_t),\label{NLE} 
\end{align}
where $\xi_t=\sigma^2/v_t^{NLE}$, $\epsilon_t^\phi=\text{Tr}\left(\bm{H}^H((\bm{H}\bm{H}^H+\xi_t\bm{I})^{-1}\bm{H}\right)/MN$ is the normalizing factor,  $\eta_t$ is a component-wise Lipschitz continuous divergence-free denoiser function of $\eta_t(\bm{r}_t)$, and $\sigma^2$ is the variance of the noise. When we perform the LE and NLE, the corresponding error variance is defined as 
\begin{eqnarray}
&&(\upsilon_t^{LE})^2=\frac{1}{MN}\mathbb{E}\{\|\bm{r}_t-\bm{s}\|^2\},\\
&&(\upsilon_t^{NLE})^2=\frac{1}{MN}\mathbb{E}\{\|\hat{\bm{s}}_t-\bm{s}\|^2\}.
\end{eqnarray}

For the NLE, it contains a denoiser and an orthogonalization. The denoiser is given by
\begin{eqnarray}
&&\bm{s}_{t+1}=\mathbb{E}\left\{\bm{s}|\bm{r}_t,\upsilon_t^{LE}\right\}\nonumber\\
&&~~~=\frac{\sum_{s_i\in\mathbb{A}}s_ie^{-\frac{\|r_i-x_i\|}{(v_t^{NLE})^2}}}{\sum_{s_i\in\mathbb{A}}e^{-\frac{\|r_i-x_i\|}{(v_t^{NLE})^2}}},
\end{eqnarray}
where $\mathbb{A}$ is the constellation set. Moreover, the NLE also guarantees the orthogonality between the input error $v_t^{LE}$ and the output error $v_{t+1}^{NLE}$. 

When channel matrix $\bm{H}$ is unitary invariant\footnote{When the channel matrix does not satisfy the unitary invariant matrix properties, we can utilize the singular value decomposition method to decompose the effective channel matrix and then use the OAMP algorithm to detect the symbols.}, the Bayesian estimator in~\eqref{LE} and the posterior mean estimator in~\eqref{NLE} are orthogonal~\cite{cheng2020integral}. Therefore, the OAMP algorithm is Bayesian optimal.
\section{Simulation Results}
 In this section, the performance of the proposed algorithm is demonstrated by the simulation results. In the simulation, we establish a carrier frequency of 5 GHz with a subcarrier spacing of 15 kHz. The number of time slots $M$ and the number of subcarriers $N$ are set as 512 and 32, respectively. The modulation alphabet is 4-QAM, and the speed of the user is 350 km/h. For the ODDM modulation, we set $a(t)$ as a square-root raised cosine pulse, where $Q=20$.
 For the DD channel, we adopt the Extend Vehicular A~(EVA) model in 3GPP~\cite{3gpp2016evolved}. 
 
\balance
\begin{figure*}[ht]
\centering
\begin{minipage}{0.3\textwidth}  \centering
  \includegraphics[width=\textwidth]{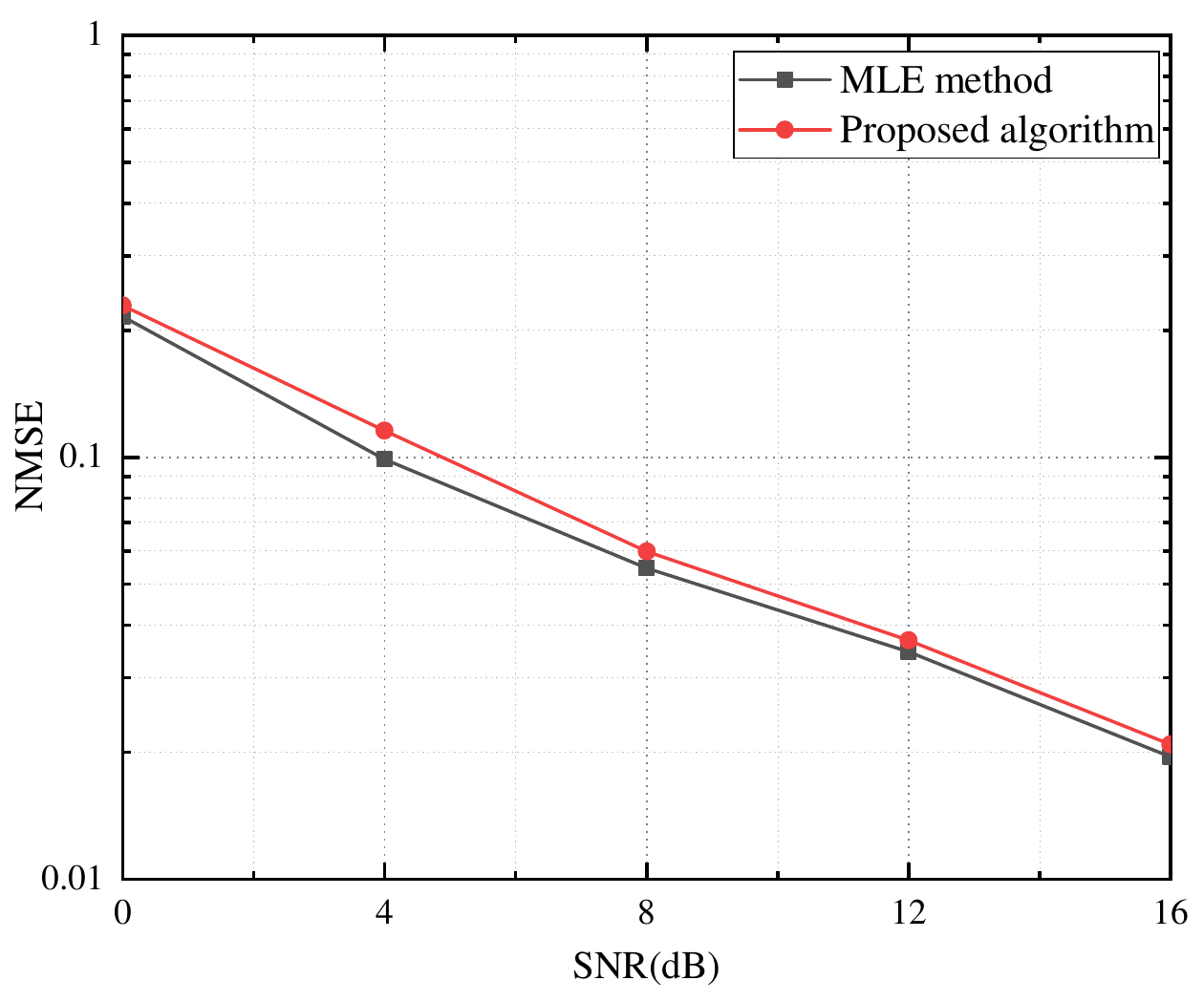}
  \caption{The NMSE performance comparison for different SNRs.}
  \label{fig:sensing_performance}
\end{minipage}
\hspace{0.02\textwidth}
\begin{minipage}{0.3\textwidth}
  \centering
  \includegraphics[width=\textwidth]{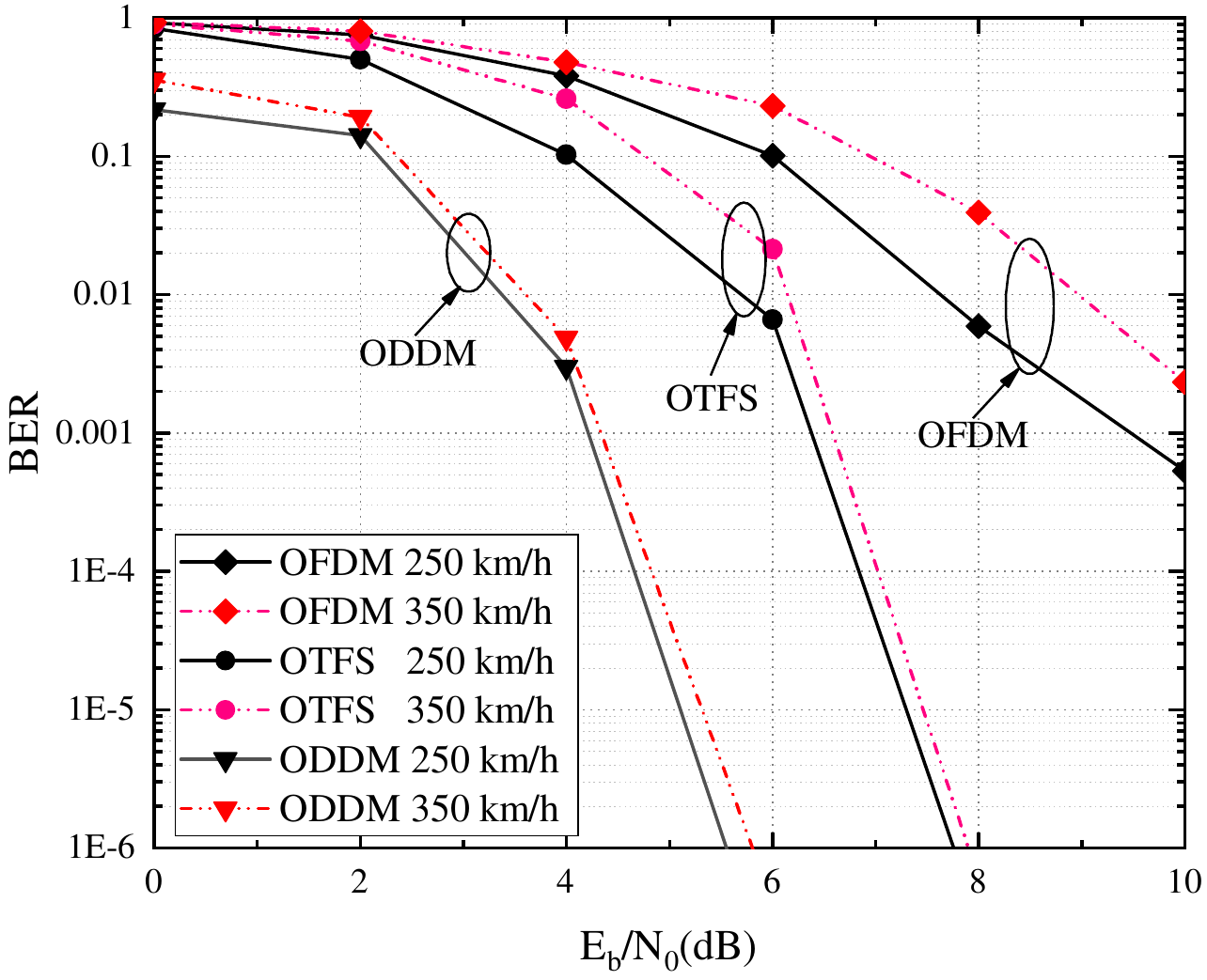}
  \caption{BER comparison, M=512, N=32, 4-QAM}
\label{fig:performance_comparsion_different_v}
\end{minipage}
\hspace{0.02\textwidth}
\begin{minipage}{0.3\textwidth}
  \centering
  \includegraphics[width=\textwidth]{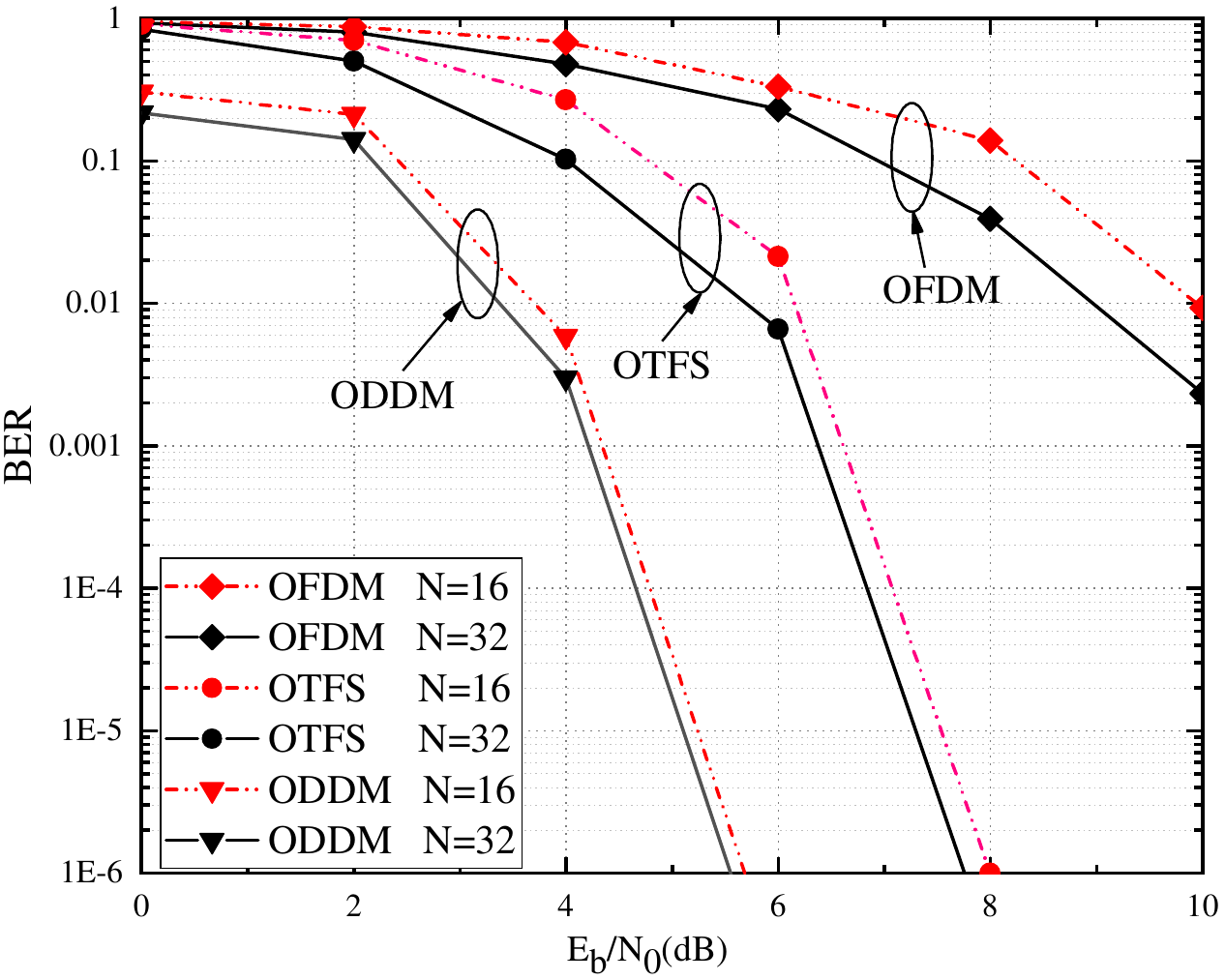}
\caption{BER comparison, M=512, 4-QAM}
\label{fig:performance_comparsion_different_N}
\end{minipage}
\end{figure*}

At first, we evaluate the sensing performance of the proposed low-complexity estimation algorithm concerning the normalized mean square error~(NMSE) in Fig.~\ref{fig:sensing_performance}. It can be seen that the NMSE performance is improved as the signal-to-noise ratio~(SNR) increases. In addition, the results estimated by the proposed algorithm are close to the MLE method, which verifies the performance of the proposed algorithm. 

Then, we compare the BER performance of the proposed algorithm with the OTFS and OFDM  modulation.  At first, we consider different mobility speed scenarios for the ODDM and OTFS algorithms in Fig.~\ref{fig:performance_comparsion_different_v}. At receiver speeds of 250 km/h and 350 km/h, the ODDM outperforms the OTFS by approximately 2.1 dB and 2.3 dB, respectively, at the BER of $10^{-6}$. The reason is that the orthogonality of the pulse $u(t)$ in ODDM modulation satisfies the perfect reconstruction condition w.r.t. the DD resolutions and the ODDM employs matched filters at the receiving end to maximize the signal-to-noise ratio. In addition, ODDM and OTFS outperform OFDM in high-mobility scenarios due to their superior BER performance against Doppler effects.


On the other hand, we consider the performance of different numbers of subcarriers in Fig.~\ref{fig:performance_comparsion_different_N}. It can be seen the ODDM modulation achieves about 2.3 dB and 2.2 dB over the OTFS when the numbers of subcarriers are 32 and 16, respectively. Similarly, the ODDM and OTFS both outperform the OFDM in high-mobility scenarios. Combined with Fig.~3, it can be concluded that the BER performance of ODDM modulation outperforms the OTFS in the ISAC system regardless of the speed of the target receiver and the number of subcarriers. Furthermore, Figs. 3 and 4 illustrate that ODDM exhibits robust performance in high-mobility scenarios across various speeds and subcarriers.

\section{Conclusion}
In this letter, we exploited the sensing parameters to enhance the communication performance in the ISAC system by utilizing the ODDM modulation, which is a pulse-train and orthogonal w.r.t. the resolutions of the DD plane. Specifically, at first, we obtained the channel information in the sensing process by designing a low-complexity algorithm to approximate the solution of the MLE scheme. After having the multi-path channel parameters, we utilized the OAMP method to detect the symbols in the communication process, where the detection scheme can achieve Bayesian optimality. Finally, we validated the detection performance of the proposed algorithm via the simulation, which outperformed the OTFS by about 2.3 dB when the BER is $10^{-6}$.



\begin{thebibliography}{1}

\bibitem{fei2023air}
Z. Fei, X. Wang, G. N. Wu, J. Huang, and J. A. Zhang, ``Air-ground
integrated sensing and communications: Opportunities and challenges,”
\emph{IEEE Commun. Mag.}, vol. 61, no. 5, pp. 55–61, May 2023.

\bibitem{hwang2008ofdm}
T. Hwang, C. Yang, G. Wu, S. Li, and G. Y. Li, ``OFDM and its wireless
applications: A survey,” \emph{IEEE Trans. Veh. Technol.}, vol. 58, no. 4, pp.
1673–1694, May 2009.

\bibitem{Zhang2023radar}
K. Zhang, W. Yuan, S. Li, F. Liu, F. Gao, P. Fan, and Y. Cai,
“Radar sensing via OTFS signaling: A delay Doppler signal processing
perspective,” in \emph{Proc. IEEE Int. Conf. Commun. (ICC)}, Rome, Italy, May 2023.

\bibitem{wang2021pilot}
S. Wang, J. Guo, X. Wang, W. Yuan, and Z. Fei, ``Pilot design and
optimization for OTFS modulation,” \emph{IEEE Wireless Commun. Lett.},
vol. 10, no. 8, pp. 1742–1746, May 2021.

\bibitem{li2023pulse}
S. Li, W. Yuan, J. Yuan, B. Bai, and G. Caire, ``On the pulse shaping
for delay-Doppler communications,” in \emph{Proc. IEEE Global Commun.
(Globecom)}, Kuala Lumpur, Malaysia, Dec. 2023.

\bibitem{lin2023multi}
H. Lin, J. Yuan, W. Yu, J. Wu, and L. Hanzo, ``Multi-carrier modulation:
An evolution from time-frequency domain to delay-doppler domain,” [Online] available:
\emph{https://arxiv.org/abs/2308.01802}, Aug. 2023.

\bibitem{lin2022orthogonal}
H. Lin and J. Yuan, ``Orthogonal delay-Doppler division multiplexing
modulation,” \emph{IEEE Trans. Wireless Commun.}, vol. 21, no. 12, pp.
11 024–11 037, Jul. 2022.



\bibitem{gan2021ris}
X. Gan, C. Zhong, C. Huang and Z. Zhang, ``RIS-Assisted Multi-User MISO Communications Exploiting Statistical CSI," \emph{IEEE Trans. Commun.}, vol. 69, no. 10, pp. 6781-6792, Oct. 2021

\bibitem{channel2021li}
L. Wei, C. Huang, G. C. Alexandropoulos, C. Yuen, Z. Zhang, and
M. Debbah, ``Channel estimation for RIS-empowered multi-user MISO
wireless communications,” \emph{IEEE Trans. Commun.}, vol. 69, no. 6, pp.
4144–4157, Jun. 2021.

\bibitem{yang2024sensing}
X. Yang, H. Li, Q. Guo, J. A. Zhang, X. Huang, and Z. Cheng,
``Sensing aided uplink transmission in OTFS ISAC with joint parameter
association, channel estimation and signal detection,” \emph{IEEE Trans. Veh.
Technol.}, vol. 99, no. 99, pp. 1–6, Jan. 2024.

\bibitem{yuan2021integrated}
W. Yuan, Z. Wei, S. Li, J. Yuan, and D. W. K. Ng, ``Integrated sensing
and communication-assisted orthogonal time frequency space transmission
for vehicular networks,” \emph{IEEE J. Sel. Topics Signal Process.}, vol. 15,
no. 6, pp. 1515–1528, Nov. 2021.

\bibitem{priya2023channel}
P. Priya, C. S. Reddy, and D. Sen, “Channel estimator and nonlinear
detector for mmWave beamformed OTFS systems in high mobility
scenarios,” \emph{IEEE Trans. Veh. Technol.}, vol. 72, no. 9, pp. 11698–11713,
Apr. 2023.

\bibitem{lin2022multicarrier}
H. Lin and J. Yuan, “Multicarrier modulation on delay-Doppler plane:
Achieving orthogonality with fine resolutions,” in \emph{Proc. IEEE Int. Conf.
Commun. (ICC)}, Rome, Italy, May 2023.

\bibitem{liu2020uplink}
Y. Liu, S. Zhang, F. Gao, J. Ma, and X. Wang, ``Uplink-aided high mobility downlink channel estimation over massive MIMO-OTFS system," \emph{IEEE J. Sel. Areas Commun.}, vol.~38, no. 9, pp. 1994--2009, Sep. 2020.

\bibitem{gaudio2020effectiveness}
L. Gaudio, M. Kobayashi, G. Caire, and G. Colavolpe, “On the effectiveness
of OTFS for joint radar parameter estimation and communication,”
\emph{IEEE Trans. Wireless Commun.}, vol. 19, no. 9, pp. 5951–5965, Sep.
2020.

\bibitem{liu2023oamp}
 L. Liu, Y. Cheng, S. Liang, J. H. Manton, and L. Ping, ``On OAMP:
impact of the orthogonal principle,” \emph{IEEE Trans. Commun.}, vol. 71,
no. 5, pp. 2992–3007, May 2023.

\bibitem{cheng2020integral}
J. Ma and L. Ping, “Orthogonal
AMP,” \emph{IEEE Access}, vol. 5, pp.2020--2033, Jan. 2017.

\bibitem{3gpp2016evolved}
3GPP, “Evolved universal terrestrial radio access (E-UTRA); user equipment
(UE) radio transmission and reception,” TS 36.101, 2016.
\end{thebibliography}
\end{document}